\documentclass[prb,twocolumn,showpacs,preprintnumbers]{revtex4-1}
\usepackage{epsfig}
\usepackage{amssymb,amsmath,amsthm,amsfonts,txfonts}
\usepackage{pstcol}
\usepackage{array}

\newcommand{\beq}{\begin{equation}}
\newcommand{\eeq}{\end{equation}}
\newcommand{\beqa} { \begin{eqnarray} }
\newcommand{\eeqa} {\end{eqnarray}}
\newcommand{\beqs} {\begin{displaymath}}
\newcommand{\eeqs} {\end{displaymath}}
\newcommand{\beqas} {\begin{eqnarray*}}
\newcommand{\eeqas} {\end{eqnarray*}}

\newcommand{\ve}[1]{{\bf #1}}  
\newcommand{\ves}[1]{\boldsymbol{#1}}
\newcommand{\tensort}[1] {\varmathbb{#1}}
\newcommand{\hydro}{{\rm hydro}}

\begin{document}
\title{A two-sphere model for bacteria swimming near solid surfaces}
\author {Jocelyn Dunstan$^1$, Gast\'on Mi\~no$^2$, Eric Clement$^2$, and Rodrigo Soto$^1$}

\affiliation{
$^1$Departamento de F\'\i sica, FCFM, Universidad de Chile, Santiago, Chile\\
$^2$PMMH-ESPCI, UMR 7636 CNRS-ESPCI-Universit\'e Paris 6 and Paris 7, 10 rue Vauquelin, 75005 Paris, France}

\date{\today}

\begin{abstract}

We present a simple model for bacteria like \emph{Escherichia coli} swimming
near solid surfaces.  It consists of two spheres of different radii
connected by a dragless rod.  The effect of the flagella is taken into
account by imposing a force on the tail sphere and opposite torques exerted
by the rod over the spheres.  The hydrodynamic forces and torques on the
spheres are computed by considering separately the interaction of a single
sphere with the surface and with the flow produced by the other sphere. 
Numerically, we solve the linear system which contains the geometrical
constraints and the force-free and torque-free conditions.  The dynamics of
this swimmer near a solid boundary is very rich, showing three different
behaviors depending on the initial conditions: (1)~swimming in circles in
contact with the wall, (2)~swimming in circles at a finite
distance from the wall, and (3)~swimming away from it.  Furthermore, the order of
magnitude of the radius of curvature for the circular motion is in the range
$8-50\,\mu$m, close to values observed experimentally.

\end{abstract}

\maketitle

\section{Introduction}
In recent decades, interest in the dynamics of self-propelled organisms such as bacteria, fish, and birds has increased enormously. This can be explained, firstly, by interest in the physics behind the phenomena, for which tools from continuum mechanics are used to describe the motion of single swimmers as well as from statistical physics to deduce collective behaviors due to their mutual interactions. Besides, study of how this locomotion affects biological processes such as human reproduction or bacterial infection remains fundamental \cite{biologicaltransport}. Finally, development of applications in which artificial swimmers could perform specific tasks in microfluidic devices is certainly an area of intense research \cite{bactratchet}.

A particular kind of self-propelled organisms are bacteria. Considering their micrometer scale and typical propulsion velocities, they are in the low-Reynolds-number regime, in which inertial effects are negligible in comparison with viscous ones. Therefore, the dynamics of the fluid is governed by the Stokes equation and immersed objects by the force-free and torque-free conditions \cite{KimKarilla}. In this category, the bacterium \emph{Escherichia coli} (\emph{E.~coli}) has been intensively studied, and much is known about its genetics, biological processes, and motility \cite{EcoliMotion}. The body cell is about 1\,$\mu$m in width and 2--5\,$\mu$m in length, and around 70\% of it is water. Its propulsion is due to rotary motors connected to around six filaments of approximately 15\,$\mu$m in length, which emerge from different points of the cell. When these filaments rotate counterclockwise, seen from behind, they form a bundle and the bacterium swims in an approximately straight line at about 20\,$\mu$m/s \cite{RandomWalk}. This is called the \emph{run} mode. If one or more filaments start to rotate in the opposite direction, the bundle (flagella) unravels, and the bacterium can change its direction, which is known as a \emph{tumble}. The alternation of these modes creates a three-dimensional random walk \cite{RandomWalk}.

In addition, when \emph{E.coli} swims near a non-slip surface, it describes clockwise circles when is viewed from above. In 1995, Frymier \emph{et al.} \cite{Frymier 1995} reported a three-dimensional tracking of the movement of two kind of \emph{E.coli}: wild type and a smooth-swimming mutants, which are unable to perform the tumbling motion described above. In the first case the bacteria describe circular segments but they leave the surface after a tumbling. Conversely, the  smooth-swimming cells swim for a much longer time along the surface. An explanation for this circular motion was given by Lauga \emph{et al.} from the fact that the head and the flagella rotate in opposite directions, which together with a stronger drag near a wall, produce a net force directed to the centre of the circle. 

An extremely interesting question is how these living entities can modify the mass, momentum, and energy transport properties in a suspension. Since the pioneering work of Wu and Libchaber \cite{Wu2000}, considerable efforts have been made to understand this issue. Recently, Mi\~no \emph{et~al.} \cite{Mino 2010} observed that, near to a solid boundary, a suspension of \emph{E.~coli} increases the diffusion coefficient of passive tracers by a factor proportional to the \emph{active flux}, i.e., the concentration of active bacteria multiplied by their mean velocity.

In the literature, there are various models for \emph{E.~coli} with diverse levels of complexity; for example, Ramia \emph{et~al.} \cite{Ramia 1993} simulated \emph{E.~coli} as a sphere joined to a single helicoidal flagellum, and the velocity field near the surface was obtained using the boundary element method. In that work, the radius of curvature was about 10\,$\mu$m, and it was reported a tendency for the bacterium to crash into the wall, i.e., it approaches the solid surface until it touches it, with the tail pointing up, at which point it stops. A different approach was made by Lauga \emph{et~al.} \cite{Lauga 2006}, who used the same bacterium geometry, but to obtain its trajectory near the surface, they calculated the resistance matrix for the swimmer. It is important to note that in their work the distance to the wall was fixed, not allowing vertical motion, which is an interesting point since the radius of curvature depends strongly on the distance to the wall. In 2008, it was reported by Li \emph{et~al.} \cite{Li 2008}, using fluorescence microscopy, that this gap is only tens of nanometers. Finally, it is important to mention the work of Gyrya \emph{et~al.} \cite{Gyrya 2010}, in which \emph{E.~coli} was modeled using two spheres joined to a point of force by a dragless rod (a similar model was introduced previously by Hernandez-Ortiz \emph{et~al.} \cite{Hernandez 2005}). There, the velocity field was obtained in the bulk for an isolated swimmer as well as for a pair of them. This latter approach inspired our work aiming to build a simple model for a bacterium using spheres, like in Gyrya {\em et~al.}, but adding the interaction with the surface and rotation of the spheres, which is necessary to reproduce circular motion near the surface. Construction of a minimal model of a bacterium that can perform this motion near a surface is the central aim of this work.

The present work is organized as follows: in Sect.~\ref{sec.twopointforce} the point-force model for bacteria is described, and the corresponding problem close to a solid boundary is discussed. Then, in Sect. \ref{sec.model}, our finite-size model is presented, obtaining its velocity in the bulk and the geometrical parameters of the model are fixed.
The problem of this swimmer near a surface is treated in Sect.~\ref{sec.hydro}, where the formalism of resistance tensors is introduced. After that, in Sect.~\ref{sec.results}, the results from numerical simulations are presented. Different types of motion are obtained, depending on the initial conditions. Finally, conclusions are presented in Sect.~\ref{sec.conclusion}.

\section{Bacterium as two point forces interacting with a solid boundary} \label{sec.twopointforce}

At large distances, a pusher bacterium such as {\em E.~coli} creates a velocity field perturbation that can be described as that produced by a force dipole. Using this model, Berke \emph{et~al.} \cite{Berke 2008} showed that a pusher swimmer experiences an attraction toward surfaces in the form of a negative vertical velocity. Moreover, the swimmer orientation is affected by the presence of the surface, adopting a stable equilibrium orientation parallel to the surface \cite{Berke 2008}.
However, the dipolar model fails to provide an appropriate description of the swimmer--surface interaction, as we show in the next paragraphs, due to the dominant finite-size effects from the swimmer. In particular, these effects regularize the apparently divergent vertical velocity obtained when the swimmer approaches the surface \cite{Berke 2008}.

\begin{figure}[htb]
\begin{center}
\includegraphics[width=0.7\linewidth]{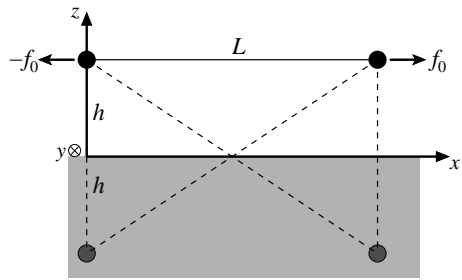}
\caption{A two-point-force model for a bacterium near a wall. The interaction with the wall is obtained using the image method proposed by Blake and Chwang \cite{Blake 1974}. The arrows represent the forces exerted by the point forces on the fluid for a pusher bacterium. Dashed lines represent the interactions due to the image forces.}
\label{monopoles}
\end{center}
\end{figure}

Using the model described in detail in Sect.~\ref{sec.model}, in which the swimmer has a finite size, it is possible to compute the limit of point forces. Considering two point forces of intensities $f_0$, separated at a distance $L$, analytic expressions can be obtained using the images system \cite{Blake 1974}. If the distance from the swimmer to the surface is $h$ (Fig.~\ref{monopoles}) and its orientation is parallel to the surface, the vertical velocity is given by
\beqa
\label{vz}
V_z &=& - \frac{3f_0 L h^3}{2\pi\eta(L^2+4h^2)^{5/2}} ,
\eeqa
where $\eta$ is the dynamic viscosity of the fluid.

For a dipole of vanishing size ($L\to0$) and fixed force dipole of strength $p=f_0L$, the vertical velocity matches the result of Berke \emph{et~al.}, which  diverges when $h\to 0$. Note that limits do not commute and the velocity divergence is only obtained when the limit in $L$ is taken before the limit in $h$. However, when $L$ is finite, the velocity is regularized for all distances and vanishing for $h\to 0$. Furthermore, the angular velocity vanishes as predicted by Berke \emph{et~al}  \cite{Berke 2008}. The factor $(L^2+4h^2)^{1/2}$ corresponds to the cross distance between one sphere and the image of the other. Therefore, the attraction to the wall is due to the two-sphere interaction through the surface. In fact, it can be easily verified that the field created by the image of a point force moving in the $x$ direction does not have a vertical velocity or angular velocity at the position of the point force itself and therefore cannot lead to an attraction toward or repulsion from the surface.

It must be remarked that the point particle approximation is not accurate, as it predicts a vanishing vertical velocity when the height of the force dipole vanishes ($h\to 0$), while the bacterium should stop its vertical motion when its body reaches the surface. Near-field corrections, obtained from the lubrication-theory approximation, should be used to describe the swimmer motion accurately. The model described in Sect.~\ref{sec.model} considers this approximation, and in Sect.~\ref{sec.results} the full swimming dynamics is described.

\section{The two-sphere model}
\label{sec.model}

\begin{figure}[htb]
\begin{center}
\includegraphics[width=0.7\linewidth]{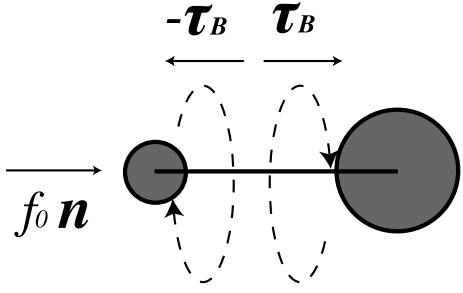}
\caption{Two-sphere model for bacteria: two spheres of different radii connected by a dragless rod, which exerts equal and opposite torques $\pm \ves \tau_B$. Over the $T$-sphere, a \emph{propulsion force} $f_0 \ve{\hat n}$ is applied by the fluid to mimic the effect of the rotating flagella.}
\label{model}
\end{center}
\end{figure}

The bacterium is modeled as two spheres, denoted by $H$ (head) and $T$ (tail), with radii $a_H$ and $a_T$, respectively, and connected by a dragless rod of length $L$ (Fig.~\ref{model}). The positions of the spheres are $\ve r_H$ and $\ve r_T$, and the unit vector along the swimmer is defined as $\hat{\ve n}=(\ve r_H-\ve r_T)/|\ve r_H-\ve r_T|=(\ve r_H-\ve r_T)/L$. To mimic the effect of the rotary motors in \emph{E.~coli}, the rod applies equal and opposite torques $\pm \ves{\tau}_B=\pm \tau_0\hat{\ve n}$ on the spheres, with positive sign for $H$. The $T$-sphere represents the bacterium flagella, and in the present model its rotation is not directly related to the propulsion; rather it is considered as an extra \emph{propulsion force}  $f_0 \ve{\hat n}$ applied by the fluid at the sphere center. The rotation of the spheres along the swimmer axis and the interaction with the solid surface are the main differences from the \emph{dumbbell} models proposed before \cite{Hernandez 2005, Gyrya 2010}.

Due to the micrometer scale of the swimmer, the total force and torque acting on each sphere are zero.
Making explicit the separation between forces and torques applied by the fluid (hydro) and those applied by the rod,
\beqa
\label{fH}
\ve F_H&=&\ve F^{\hydro}_H+\ve F^{\rm int}=0 ,\\
\label{fT}
\ve F_T&=&\ve F^{\hydro}_T-\ve F^{\rm int}=0 ,\\
\label{tH}
\ves \tau_H&=&\ves \tau^{\hydro}_H+\ves \tau_{B}=0 ,\\
\label{tT}
\ves \tau_T&=&\ves \tau^{\hydro}_T-\ves \tau_{B}=0 ,
\eeqa
where $\ve F^{\rm int}$ is the internal force exerted by the rod.
The hydrodynamic forces and torques are those over smooth spheres with the nonslip boundary condition (for example the Stokes drag in the case of spheres in the bulk). The exception is the $T$-sphere where, to mimic the propulsion effect of the rotating flagella, an additional propulsion force $f_0\ve{\hat n}$ is added to $\ve F^{\hydro}_H$.
From Eqs. (\ref{tH}) and (\ref{tT}),
\beq
(\ves \tau_H-\ves \tau_T) \cdot \ve{\hat n}=(\ves \tau^{\hydro}_H-\ves \tau^{\hydro}_T)\cdot \ve{\hat n}+2 \ves \tau_B \cdot \ve{\hat n}=0 .
\eeq
The torque intensity $\tau_0=\ves \tau_B \cdot \ve{\hat n}$ has been measured for \emph{E.~coli}, obtaining values of the order of 1\,pN\,$\mu$m
($\tau_0=1.26$\,pN\,$\mu$m in Ref. \cite{Reid 1993}, value that we use in the present article).

From the geometrical constraints of the swimmer, the angular velocities of the spheres, perpendicular to the swimmer axis, must be equal (denoted by $\ves \Omega_B$), and since the rod is rigid in these directions,
\beq
(\ves \Omega_H-\ves \Omega_T) \times \hat{\ve n}=\ve 0 .
\eeq
Besides, the linear velocities of the spheres are related by
\beq
\ve U_H=\ve U_T+ L \ve{\hat n} \times \ves \Omega_B ,
\eeq
where $\ve U_H$ ($\ve U_T$) and $\ves \Omega_H$ ($\ves \Omega_T$) are the linear and angular velocity of the sphere $H$ ($T$).

\subsection{Swimmer in the bulk}

In the bulk, neglecting the Fax\'en corrections~\cite{KimKarilla}, the hydrodynamic forces and torques on the spheres 
 are given by
\beqa
\ve F_H^{\hydro}&=&-6 \pi \eta a_H \left[\ve U_H-\ve u^H(\ve r_H)\right] ,\label{fHhydro}\\
\ve F_T^{\hydro}&=&-6 \pi \eta a_T \left[\ve U_T-\ve u^T(\ve r_T)\right]+f_0 \hat{\ve n} ,\label{fThydro}\\
\ves \tau_H^{\hydro}&=&-8 \pi \eta a_H^3 \left[\ves \Omega_H-\frac{1}{2}\nabla\times \ve u^H(\ve r_H)\right] ,\label{tauHhydro}\\
\ves \tau_T^{\hydro}&=&-8 \pi \eta a_T^3 \left[\ves \Omega_T-\frac{1}{2}\nabla\times \ve u^T(\ve r_T)\right].\label{tauThydro}
\eeqa
These expressions consider the far-field approximation, that is, when the  separation between the spheres is large in comparison with the radius. The neglected contributions are of order ${\cal O}(a^2/L^2)$. Here $\ve u^{H}(\ve r_H)$ is the velocity of the fluid at the position of the center of sphere $H$ but calculated in its absence (the same being valid for sphere $T$). Being more specific, if the swimmer is isolated, the field produced by the two spheres in the far-field approximation is
\beq
\ve u(\ve r)=-\tensort{G}(\ve r-\ve r_H) \cdot \ve F^{\hydro}_H-\tensort{G}(\ve r-\ve r_T) \cdot \ve F^{\hydro}_T ,
\eeq
where $\tensort{G}(\ve r)$ is the Oseen tensor defined as $\tensort{G}(\ve r)=\frac{1}{8 \pi \eta |\ve r|} \left(\mathbb{I}+\frac{\ve r \ve r^T}{|\ve r|^2}\right)$ with $\mathbb{I}$ the unit tensor. Again, the Fax\'en corrections ${\cal O}(a^2/L^2)$ haven been neglected.
Note that $\ve F^{\hydro}$ are forces applied by the fluid. Hence, the fields $\ve u^H (\ve r)$ and $\ve u^T (\ve r)$ are
\beqa
\ve u^H(\ve r)=\ve u (\ve r)+\tensort{G}(\ve r-\ve r_H) \cdot \ve F^{\hydro}_H , \label{eq.uH}\\
\ve u^T(\ve r)=\ve u (\ve r)+\tensort{G}(\ve r-\ve r_T) \cdot \ve F^{\hydro}_T \label{eq.uT}.
\eeqa

To obtain the velocity of the swimmer in the bulk, like in Ref.~\cite{Gyrya 2010}, we replace Eqs. (\ref{eq.uH}) and (\ref{eq.uT})  in (\ref{fHhydro}--\ref{tauThydro}), and with the zero-force and zero-torque conditions together with the geometrical constraints, we obtain
\beqa
\ve U_H&=&\ve U_T=\frac{2L-3 a_H}{12 \pi \eta[L(a_H+a_T)-3 a_H a_T]} \, f_0 \hat{\ve n}=U_0 \hat{\ve n}, \label{bulkV}\\
\ves \Omega_H&=&\frac{L^3-a_H^3}{8 \pi \eta a_H^3 L^3}\,  \tau_0 \ve{\hat n} ,\label{bulkWH}\\
\ves \Omega_T&=&\frac{a_T^3-L^3}{8 \pi \eta a_T^3 L^3} \, \tau_0 \ve{\hat n} .\label{bulkWT}
\eeqa

Thus, the linear velocity is proportional to the propulsion force, with a constant of proportionality depending on the swimmer geometry. Similarly, the angular velocities of the spheres are along the swimmer axis (i.e., it swims in a straight line), with magnitude proportional to the imposed torque along the axis, depending also on the relation between the geometrical parameters $a_H$, $a_T$, and $L$. Since the torques applied by the rod have opposite sign, the $H$-sphere rotates in clockwise direction while the $T$-sphere rotates counterclockwise, both as seen from behind the swimmer.

\subsection{Choice of geometrical parameters} \label{geom.param}
The geometrical parameters of the model $a_H$, $a_T$, and $L$ were chosen in the following way: the head sphere was related to the cell body using the \emph{equivalent sphere} concept~\cite{Lauga 2006,Happel 1965}. Following Lauga {\em et~al.}, an equivalent sphere having the same viscous resistance as a prolate ellipsoid moving along its axis of symmetry has radius
\beq
a=\frac{4 w}{3} \left(\frac{2 \phi^2-1}{(\phi^2-1)^{3/2}}\ln \left(\frac{\phi+\sqrt{\phi^2-1}}{\phi-\sqrt{\phi^2-1}} \right)-\frac{2 \phi}{\phi^2-1}\right)^{-1} ,
\label{REq}
\eeq
where $w$ is the width and $\phi$ is the aspect ratio of the ellipsoid. In the case of \emph{E.~coli}, the radius of the equivalent sphere varies from 0.89 to 1.07\,$\mu$m \cite{Lauga 2006}. We therefore fix $a_H=1\, \mu$m.

Recent experiments have shown that, when the flow field around an \emph{E.~coli} bacterium is measured, the separation of the associated force dipole is on the order of 2\,$\mu$m \cite{Jorn}. With this information, we fix the separation between the spheres at $L=2 \, \mu$m.

We took the radius of the tail sphere as $a_T=0.5 \, \mu$m. This choice is rather arbitrary, since this model is a great simplification of \emph{E.~coli} and a direct comparison is not possible. Other values give qualitatively similar results.

Finally, equation (\ref{bulkV}) allows one to fix $f_0$ for a given geometry by knowing the experimental value obtained in Ref.~\cite{RandomWalk}, $U_0=20 \, \mu$m/s. Using the values above, $f_0=1.13$\,pN.

\section{Two-sphere swimmer in the presence of a solid boundary}
\label{sec.hydro}

When the swimmer is near to a solid boundary, there is a complex hydrodynamic interaction with the wall. One possible approach to calculate this interaction is to use the image method proposed by Blake and Chwang \cite{Blake 1974}, i.e., to consider the bacterium as two equal and opposite point forces and point torques. Since this is a far-field approximation, and the swimmer can be very close to (namely touching) the wall, we will introduce corrections given by the Fax\'en formula \cite{Swan 2007} and lubrication theory \cite{Goldman 1967}.

Using the resistance matrix formalism, which is based on the linearity of the Stokes equations, the hydrodynamic forces and torques on the spheres that compose the swimmer are given by the relation
\begin{equation}
 \left( \begin{array}{c} \ve F^{\hydro}_H \\ \ves \tau^{\hydro}_H \\ \hline\ve F_T^{\hydro} \\ \ves \tau_T^{\hydro} \end{array}\right)=
 \left(  \begin{array}{c | c}
 \parbox[c][0.8cm][c]{0.6cm}{$\mathcal{R}_H$} &
 \parbox[c][0.8cm][c]{0.6cm}{$\mathcal{R}_{HT}$}  \\
 \hline
 \parbox[c][0.8cm][c]{0.6cm}{$\mathcal{R}_{TH}$} &
 \parbox[c][0.8cm][c]{0.6cm}{$\mathcal{R}_T$}  \end{array}\right)
 \left( \begin{array}{c} \ve U_H \\ \ves \Omega_H \\ \hline \ve U_T \\ \ves \Omega_T \end{array}\right)
+ \left( \begin{array}{c} 0\\ 0\\ \hline f_0 \hat{\ve n}  \\0  \end{array}\right) ,
 \label{system}
\end{equation}
where the $6\times6$ matrices $\mathcal{R}_H$ and $\mathcal{R}_T$ give the interaction of each sphere with the wall, and $\mathcal{R}_{HT}$ and $\mathcal{R}_{TH}$ are the couplings between them considering the wall. 
These matrices already take into account the Fax\'en and lubrication corrections.
In the article of Swan and Brady \cite{Swan 2007}, expressions for these matrices are given in the case of spheres of the same radius in the far-field approximation; the generalization to spheres of different radii is rather complicated. This approach, however, is too rigid, because if two or more swimmers are considered, the size of the resistance matrix grows accordingly, and even more so the complication of the computation of its elements.

As a simplification to this approach, we compute the interaction of the swimmer with the wall as the complete interaction of each sphere with the boundary (considering the far- and near-field contributions), while the coupling between the spheres is obtained from the interaction with the \emph{external flow} produced by the other sphere. For example, for the $H$-sphere,
\begin{equation}
\left[ \begin{array}{c} \ve F^{\hydro}_H \\ \ves \tau^{\hydro}_H  \end{array}\right] =\left[ \begin{array}{c c} \mathcal{R}^{FU}_H & \mathcal{R}^{F \Omega}_H  \\ \mathcal{R}^{\tau U}_H & \mathcal{R}^{\tau \Omega}_H \end{array}\right] \left[ \begin{array}{c} \ve U_H \\ \ves \Omega_H  \end{array}\right]+ \left[ \begin{array}{c c} \mathcal{R}^{FU^\infty}_H & \mathcal{R}^{F \Omega^\infty}_H  \\ \mathcal{R}^{\tau U^\infty}_H & \mathcal{R}^{\tau \Omega^\infty}_H \end{array}\right] \left[ \begin{array}{c} \ve U^\infty_H \\ \ves \Omega^\infty_H  \end{array}\right] ,
\label{separacion}
\end{equation}
where the matrix elements $\mathcal{R}$ are the $3\times3$ resistance tensors, and $(\ve U_H^\infty, \ves \Omega^\infty_H)$ correspond to the fluid velocity and fluid angular velocity at the position of the $H$-sphere due to the presence of the $T$-sphere. This approach has the benefit of being simple enough to extend to the case of many swimmers or other passive objects, or to include an imposed flow. The associated velocity fields are introduced in the $(\ve U^\infty, \ves \Omega^\infty)$ terms. This approximation is valid when the two spheres are sufficiently separated and no close field interactions are present. As a consequence, the corrections are of order ${\cal O}(a^2/L^2)$.

The hydrodynamic coupling between the spheres is obtained by computing the flow produced by each of them and evaluating it at the position of the other sphere. In the presence of a plane wall, the Oseen tensor is modified by image terms. The image set of a force monopole is not simply an opposite force as in electrostatics, but also includes a force quadrupole and a mass source dipole; the combination of these together with the original force monopole produces the flow that satisfies the nonslip boundary condition at the wall \cite{Blake 1974}. In the case of a sphere close to the plane, the flow should be modified to include the Fax\'en correction of the spheres and the lubrication flows. In the present model we will neglect the Fax\'en corrections, as they contribute subdominant terms. Thus, the flow produced by the $H$-sphere (that applies a force $\ve F$ and a torque $\ves \tau$ over the fluid), evaluated at $\ve r_1=(x_1,y_1,z_1)$, is
\beqa
[\ve u (\ve r_1)]_i&=&\frac{F_j}{8 \pi \eta} \left[ \left( \frac{\delta_{ij}}{r}+\frac{r_ir_j}{r^3}\right)-\left(\frac{\delta_{ij}}{\bar r}+\frac{\bar r_i\bar r_j}{\bar r^3}\right) \right. \nonumber\\
&&\left. +2h(\delta_{j1}\delta_{1 k}+\delta_{j2}\delta_{2 k}-\delta_{j3}\delta_{3k} )\frac{\partial}{\partial \bar r_k}\left\{ \frac{h \bar r_i}{\bar r^3}-\left(\frac{\delta_{i3}}{\bar r}+\frac{\bar r_i \bar r_3}{\bar r^3} \right)\right\}\right] \nonumber\\
&& +\frac{1}{8 \pi \eta} \left[ \frac{\varepsilon_{ijk} \tau_j r_k}{r^3}-\frac{\varepsilon_{ijk} \tau_j \bar r_k}{\bar r^3}+2h \varepsilon_{kj3} \tau_j \left( \frac{\delta_{ik}}{\bar r^3}-\frac{3 \bar r_i \bar r_k}{\bar r^5}\right) \right. \nonumber\\
&&\left. + 6 \varepsilon_{kj3} \frac{\tau_j \bar r_i \bar r_k \bar r_3}{\bar r^5} \right] ,
\label{stokesletRotletPared}
\eeqa
where $\ve y=\{y_1,y_2,h\}$ is the position of the force point, $\ve r=\{x_1-y_1, x_2-y_2,x_3-h\}$ is the vector from the singularity to the observation point, and $\ve{\bar r}=\{x_1-y_1, x_2-y_2,x_3+h\}$ is the vector from the position of the image to the observation point.
In the coordinate axes, $x$ and $y$ are the planar directions and $z$ is the perpendicular direction to the plane, as shown in Fig.~\ref{monopoles}.

Considering the symmetries and the linearity of the flow at low Reynolds number, the resistance  tensors have the following general expressions for a single sphere: 
\begin{align}
\mathcal{R}^{FU} &= \left[ \begin{array}{ccc}
R^{FU}_{\parallel} & 0 & 0\\
0 & R^{FU}_{\parallel} & 0\\
0 & 0 &R^{FU}_{\perp}\\
\end{array}\right] ,\\
\mathcal{R}^{\tau\Omega} &= \left[ \begin{array}{ccc}
R^{\tau\Omega}_{\parallel} & 0 & 0\\
0 & R^{\tau\Omega}_{\parallel} & 0\\
0 & 0 &R^{\tau\Omega}_{\perp}\\
\end{array}\right] ,\\
\mathcal{R}^{\tau U} &= \left(\mathcal{R}^{F\Omega}\right)^T = \left[ \begin{array}{ccc}
0 & R^{\tau U}_{xy} & 0\\
-R^{\tau U}_{xy} & 0 & 0\\
0 & 0 &
\end{array}\right],
\end{align}
and similarly for the tensors that couple with the velocity field, $\mathcal{R}^{FU^\infty}$, $\mathcal{R}^{F \Omega^\infty}$, $\mathcal{R}^{\tau U^\infty}$, and $\mathcal{R}^{\tau \Omega^\infty}$.

In the bulk, the resistance tensors take the asymptotic values
\begin{align*}
&\mathcal{R}^{FU}=-\mathcal{R}^{FU^\infty}=-6 \pi \eta a\,  \mathbb{I},  \quad
\mathcal{R}^{\tau \Omega}=-\mathcal{R}^{\tau \Omega^\infty}=-8 \pi \eta a^3\, \mathbb{I}, \quad\\
&\mathcal{R}^{\tau U}=\mathcal{R}^{\tau U^\infty}=\mathcal{R}^{F \Omega}=\mathcal{R}^{F \Omega^\infty}=0 ,
\end{align*}
where $a$ is the radius of the corresponding sphere. Together with the geometrical constraints, these produce the bulk results (\ref{bulkV}), (\ref{bulkWH}), and (\ref{bulkWT}).

To obtain each term in the resistance tensors for a sphere in the presence of a planar wall, we take advantage of the linearity of the Stokes equation. In fact, it is possible to consider separately the cases of a sphere moving and rotating in a quiescent fluid and that of a stationary sphere in an ambient fluid. The first case allows $\mathcal{R}^{FU}$, $\mathcal{R}^{F \Omega}$, $\mathcal{R}^{\tau U}$, and $\mathcal{R}^{\tau \Omega}$ to be computed, and the second case is used to obtain $\mathcal{R}^{FU^\infty}$, $\mathcal{R}^{F \Omega^\infty}$, $\mathcal{R}^{\tau U^\infty}$, and $\mathcal{R}^{\tau \Omega^\infty}$.

\subsection{Translational and rotational motion of a sphere in the presence of a plane wall in a fluid at rest at infinity}

\begin{figure}[htb]
\begin{center}
\includegraphics[width=0.5\linewidth]{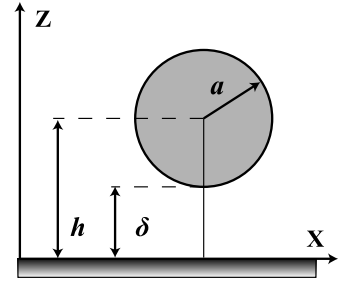}
\caption{Sketch of the quantities used to compute the resistance tensors. The surface is in the $x$--$y$ plane, the sphere radius is $a$, $h$ is the distance from the sphere center to the plate, and $\delta=(h-a)$ is the gap.}
\label{def}
\end{center}
\end{figure}

In the far-field limit, the resistance for a sphere of radius $a$ translating parallel and perpendicular to a wall, with its center located at distance $h$ from the wall (Fig.~\ref{def}), is \cite{Swan 2007}
\beqa
\nonumber
\frac{\mathcal{R}_{\parallel}^{F U}}{6 \pi \eta a}\Big|_{h \gg a}&=&- \left[1-\frac{9}{16} \left(\frac{a}{h} \right)+\frac{1}{8} \left(\frac{a}{h} \right)^3\right]^{-1} ,\\
\nonumber
\frac{\mathcal{R}_{\perp}^{F U}}{6 \pi \eta a}\Big|_{h \gg a}&=&- \left[1-\frac{9}{8} \left(\frac{a}{h} \right)+\frac{1}{2} \left(\frac{a}{h} \right)^3\right]^{-1} .\\
\eeqa
Furthermore, due to the presence of the solid boundary, there is a coupling between the translation and the torque
\beqa
\frac{\mathcal{R}_{xy}^{F \Omega}}{6 \pi \eta a^2}\Big|_{h \gg a}&=&-\frac{\mathcal{R}_{yx}^{F \Omega}}{6 \pi \eta a^2}\Big|_{h \gg a} =\frac{1}{8} \left(\frac{a}{h} \right)^4 ,
\eeqa
where $\mathcal{R}^{F\Omega}=[\mathcal{R}^{\tau U}]^T$ (see Ref. \cite{KimKarilla} for a description of the general properties of the resistance matrix). Similarly, the resistance experienced by a sphere rotating parallel and perpendicular to a wall in the far-field approximation is \cite{Swan 2007}
\beqa
\nonumber
\frac{\mathcal{R}_{\parallel}^{\tau \Omega}}{8 \pi \eta a^3}\Big|_{h \gg a}&=&-\left[1-\frac{5}{16} \left(\frac{a}{h} \right)^3\right]^{-1} ,\\
\label{RTW}
\frac{\mathcal{R}_{\perp}^{\tau \Omega}}{8 \pi \eta a^3}\Big|_{h \gg a}&=&-\left[1-\frac{1}{8} \left(\frac{a}{h} \right)^3\right]^{-1} .
\eeqa

On the other hand, when the sphere is close to the plane [that is, when the gap distance is much smaller than the sphere size, $\delta=(h-a) \ll a$], the asymptotic lubrication theory must be used. Goldman \emph{et~al.} \cite{Goldman 1967} obtained the following asymptotic expressions:
\beqa
\nonumber
\frac{\mathcal{R}_{\parallel}^{F U}}{6 \pi \eta a}\Big|_{h \rightarrow a}&=&\frac{8}{15} \ln \left(\frac{\delta}{a} \right)-0.9588 ,\\
\nonumber
\frac{\mathcal{R}^{F \Omega}_{xy}}{6 \pi \eta a^2}\Big|_{h \rightarrow a}&=&-\frac{2}{15} \ln \left(\frac{\delta}{a} \right)-0.2526 ,\\
\frac{\mathcal{R}_{\parallel}^{\tau \Omega}}{8 \pi \eta a^3}\Big|_{h \rightarrow a}&=&\frac{2}{5} \ln \left(\frac{\delta}{a} \right)-0.3817 .
\eeqa

We propose global interpolation formulas, valid in both limits, with the following generic form
\beq
\left[a_0 +b_0 \, x \right]  \log \left( \frac{x-1}{x} \right)+c_0 +d_0 \frac{1}{x}+e_0 \frac{1}{x^2} +f_0 \frac{1}{x^3}  ,
\label{int}
\eeq
where $x \equiv h/a$ is the normalized distance from the center of the sphere. The coefficients of the interpolation are fixed to match the two asymptotic  expressions up to the known orders in each case. For example, for $\mathcal{R}^{FU}_\parallel$, the expression is
\beqa
\frac{\mathcal{R}_{\parallel}^{F U}}{6 \pi \eta a} &=& \left(1.5296 -0.9963 \frac{h}{a} \right)  \log \left( \frac{h-a}{h} \right)-1.9963 , \nonumber \\
&& +0.4689 \left(\frac{a}{h}\right) +0.4327 \left(\frac{a}{h}\right)^2 +0.1358 \left(\frac{a}{h}\right)^3 .\label{RFUpar}
\eeqa
The expressions of the global fits in the other cases are given in Appendix \ref{appendix.fitsR}.

On the other hand, for a sphere moving perpendicular to the wall, we use the resistance obtained by Bevan and Prieve \cite{Bevan 2000} as a regression of the exact solution given by Brenner \cite{Brenner 1961}
\beq
\frac{\mathcal{R}_{\perp}^{F U}}{6 \pi \eta a}=-\frac{2+9\left(\frac{\delta}{a} \right)+6\left(\frac{\delta}{a} \right)^2}{2 \left(\frac{\delta}{a} \right)+6\left(\frac{\delta}{a} \right)^2} .\label{RFUperp}
\eeq
It is remarkable that this expression is valid for both short and long distances from the wall. Finally, rotation perpendicular to the wall has been rarely treated in the literature; in the work of Lee and Leal \cite{Lee 1980}, the Stokes equation is solved for a sphere moving and translating in the presence of a wall. The expressions for the resistances are given in terms of infinite series. In particular,
\beqa
\frac{\mathcal{R}_{\perp}^{\tau \Omega}}{8 \pi \eta a^3}&=&-\frac{\sinh^2 \eta_0}{\sqrt{2}}\sum_{n=1}^{\infty} n(n+1)\frac{2 \sqrt{2} c e^{(n+1/2) \eta_0}}{\sinh[(n+1/2) \eta_0]} ,
\label{Rperp}
\eeqa
where $\eta_0=-{\rm arccosh}(x) $ and $c=\sqrt{x^2-1}$, with $x=h/a$. For small gaps, this expression goes as $\mathcal{R}_{\perp}^{\tau \Omega} = {8 \pi \eta a^3}[ 1.202 + 0.414 \, (x-1)\, \log (x-1)]$.
A global interpolation between this limit and the expression (\ref{RTW}) is proposed in the form
\beq
\left[a_0 (x-1) +b_0 \frac{(x-1)^2}{x}\right]  \log \left( \frac{x-1}{x} \right)+c_0 +d_0 \frac{1}{x}+e_0 \frac{1}{x^2} +f_0 \frac{1}{x^3} ,
\label{int2}
\eeq
and the coefficients are given in Appendix \ref{appendix.fitsR}.

\subsection{Immobilized sphere in an ambient flow}

In the work of Goldman \emph{et~al.} that considers a shear flow passing around an immobilized sphere \cite{Goldman 1967}, it was found that the shear flow induces forces and torques. The associated resistances approach finite values when the sphere is in contact with the wall:
\beq
\frac{\mathcal{R}^{F U^\infty}_{\parallel}}{6 \pi \eta a}\Big|_{h \rightarrow a}=1.7005 \hspace{1cm}  \frac{\mathcal{R}^{\tau U^\infty}_{\parallel}}{8 \pi \eta a^3}\Big|_{h \rightarrow a}=0.9440 .
\eeq
As in the case of a sphere in the bulk, for large distances, the resistances are the same as for a sphere translating and rotating in a quiescent fluid but with  opposite sign,
\beq
\frac{\mathcal{R}^{F U^\infty}_{\parallel}}{6 \pi \eta a}\Big|_{h \gg a}=1+\frac{9}{16}\left(\frac{a}{h} \right) \hspace{1cm}  \frac{\mathcal{R}^{\tau U^\infty}_\parallel}{8 \pi \eta a^3}\Big|_{h \gg a}=1-\frac{3}{16}\left(\frac{a}{h} \right) .
\eeq
The global interpolation between both limits is given in Appendix A.

On the other hand, when the flux is perpendicular to the wall, there is no known expression for the lubrication approximation. For simplicity, we assume that, as for $\mathcal{R}^{F U^\infty}_{\parallel}$ and $\mathcal{R}^{\tau U^\infty}_{\parallel}$, this resistance does not diverge when $\delta\to0$. Therefore, we use the expression for large distances of a sphere moving perpendicular in a rest fluid, with the opposite sign, i.e.,
\beq
\frac{\mathcal{R}^{F U^\infty}_{\perp}}{6 \pi \eta a}\Big|_{h \gg a}=1+\frac{9}{8}\left(\frac{a}{h} \right) \hspace{1cm}  \frac{\mathcal{R}^{\tau \Omega^\infty}_\perp}{8 \pi \eta a^3}\Big|_{h \gg a}=1+\frac{1}{8}\left(\frac{a}{h} \right)^3 .
\eeq

Finally, there remains the coupling between force and rotation (and equivalently torque and translation) due to this external flow. In Ref. \cite{Goldman 1967} these coupling were not considered, because the imposed  flow was a simple linear shear flow characterized by a single parameter, the shear rate. Hence the local velocity $\ve U^\infty$ is proportional to the local angular velocity $\ve\Omega^\infty$. We have extended this to consider external flows that may not be linear and that are characterized by the local velocity and angular velocity, independently. A condition for consistency is that, in the case of linear flows, the known result should be recovered. For simplicity, and without any other information, we neglect these couplings.

\section{Results}
\label{sec.results}

\subsection{Stable swimming in contact with the surface} \label{sec.regularization}
At each time step, the linear system of equations for the linear and angular velocities is solved numerically. The associated matrix depends on the orientation and vertical position of the swimmer, becoming singular when any of the spheres touches the wall. The swimmer motion, that is its position and orientation as a function of time,  is obtained using a fourth order Runge-Kutta method with adaptive time step.     

For a wide range of geometrical parameters ($0.5\,\mu$m$<a_H<2\,\mu$m, $0.2\,\mu$m$<a_T<1\,\mu$m, and $1\,\mu$m$<L<4\,\mu$m), including those in Sect.~\ref{geom.param}, the dynamics of our model is as follows.
If the swimmer is initially pointing toward the surface, it approaches the surface in almost a straight line until it is very close to it, at which point it slows down and orients such that both spheres almost touch the surface (close but not exactly parallel to the surface because of the difference in radii). Then, the complete swimmer is attracted to the surface until it touches it in finite time. At this stage, the swimmer stops any motion (translation and rotation) in a stable state except for eventual numerical errors if the time step is not small enough. This motion is similar to that previously reported for other models: when the initial conditions of the swimmer are prepared such that it reaches the surface, it crashes into the surface and stops moving \cite{Ramia 1993}. In our model, the swimmer stops with both spheres touching the surface. 

The stopping process can be understood because the resistance tensors diverge when the spheres touch the surface. Therefore, the finite propulsion force and rod torque cannot produce motion. However, the nature and character of the divergence of the different tensors is different, requiring deeper analysis. The vertical resistance $\mathcal{R}^{FU}_\perp$ (\ref{RFUperp}) diverges strongly as $1/\delta$. This divergence is necessary to avoid penetration of the body into the surface and it is always present, quite independently of the precise geometrical shape of the swimmer. On the other hand, the parallel resistance  $\mathcal{R}^{FU}_\parallel$ (\ref{RFUpar}) has a soft divergence as $\log(\delta)$, which is only noticeable when the sphere is extremely close to (namely, in contact with) the surface. However, before becoming divergent, in real swimmers, the roughness at the nanoscale regularizes this resistance \cite{Rough}. The same happens for all resistances with the exception of $\mathcal{R}^{FU}_\perp$. Therefore the resistances are naturally regularized at the nanometer scale, and eventually electrochemical forces enter into play, also avoiding that the spheres touch the surface \cite{Li 2008}.

To keep the model at a hydrodynamic level, without including extra forces, the vertical resistance   $\mathcal{R}^{FU}_\perp$ is modified such that it diverges at a small but finite value of $\delta$; therefore, the vertical movement is stopped before the other resistances diverge. Specifically, instead of (\ref{RFUperp}), we take
\beq
\frac{\mathcal{R}_{\perp}^{F U}}{6 \pi \eta a}=-\frac{2+9\left(\frac{\delta^*}{a} \right)+6\left(\frac{\delta^*}{a} \right)^2}{2 \left(\frac{\delta^*}{a} \right)+6\left(\frac{\delta^*}{a} \right)^2} ,\label{RFUperpNew}
\eeq
where $\delta^*=(\delta-\epsilon)$ and the gap size is chosen as $\epsilon=0.01 a_H=10$\,nm. This value agrees with the swimming distance of some tens of nanometers measured by Li \emph{et~al.}, which is consistent with the prediction of the stable position including the Derjaguin--Landau--Verwey--Overbeek (DLVO) potential \cite{Li 2008}.

With this regularization scheme, the swimmer can perform stable motion in contact with the surface, as described in the next section.

\subsection{Swimming modes for fixed parameters}
\begin{figure*}[htb]
\begin{center}
\includegraphics[width=\linewidth]{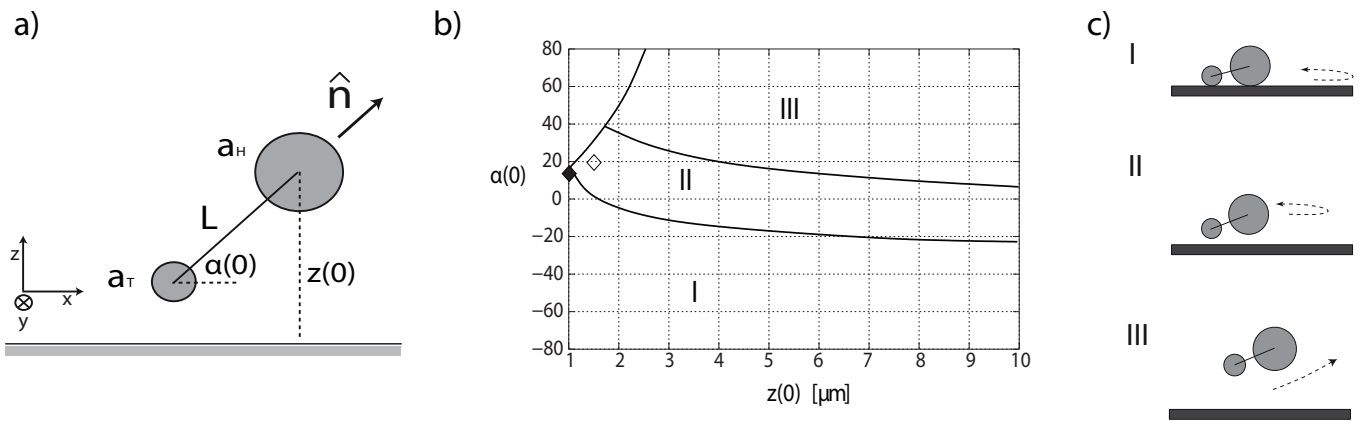}
\caption{(a) Initial condition for the swimmer: the orientational angle with respect to the surface is $\alpha(0)$, and the position of the center of sphere $H$ is $z(0)$. The geometrical parameters are $a_H=1\, \mu$m, $a_T=0.5\, \mu$m, $L=2\, \mu$m, and the gas size is set to $\epsilon=0.01\, \mu$m . (b) Phase diagram of the initial conditions. Three situations are observed: (I)~swimming in circles in contact with the wall, (II)~swimming parallel to the wall at a finite distance, and (III)~swimming away from the wall. The final states for regimes~I and II are indicated by a symbol. The upper left corner is forbidden by the condition that the $T$ sphere must be above the surface. (c) Sketch of the three final regimes.}
\label{PD}
\end{center}
\end{figure*}

 Depending on the initial condition, characterized by the initial height of the sphere $H$, $z(0)$, and the initial orientation with respect to the surface $\alpha(0)$ (Fig.~\ref{PD}a), the swimmer
shows different asymptotic behaviors.
Three different behaviors were identified: (I)~swimming in circles in contact with the plate, (II)~swimming parallel to the plate at a finite distance, and (III)~swimming away from the surface. Figure~\ref{PD}b presents a phase diagram of the initial conditions that lead to these final regimes, which are schematically presented in Fig.~\ref{PD}c. Animations of these three regimes can be seen in the Supplementary Material \cite{animations}.

\subsubsection{Swimming in circles touching the surface}

When the swimmer is initially close to the surface and pointing towards it (negative angle $\alpha$), it approaches the surface, orients such that both spheres almost touch the surface, and swims in circles. This state is stable, and the swimmer remains in this state almost in contact with the surface thanks to the regularization described in Sect.~\ref{sec.regularization}.
All initial conditions in region~I of the phase diagram end in the same final state, characterized by $z_H-R_H=z_T-R_T=\epsilon$. Consequently, the final orientation, which depends on the swimmer geometry, is given by $\alpha=14^\circ$. The swimmer velocity is $V_{\rm circle}=9.8\, \mu$m/s, and the radius of curvature  $R_C=15.8\, \mu$m. Experiments show a wide dispersion of velocities and radius of curvature of {\em E.~coli} swimming in contact with a solid surface \cite{tobepublished}. The computed values for this model fall within the observed experimental ranges.

\subsubsection{Swimming  in circles  at a finite distance}

When the initial orientation angles are higher than in the former regime, while still not being too large, the swimmer ends up in a stable state, moving parallel to the surface at a finite distance. The final configuration is given by $z_H=1.51\, \mu$m, $z_T= 0.84\, \mu$m, and orientation with respect to the surface $\alpha=19^\circ$. Due to the distance to the wall, the hydrodynamic interaction with it is  strong, and the radius of curvature of this state is $R_C=12.9\,\mu$m while the swimming velocity is that of the bulk $V_{\rm parallel}=20\,\mu$m/s. All the initial conditions in region~II of the phase diagram end in the same state. This regime has not been identified in experiments as a different regime from the previous one. Possibly because the radii of curvature are similar and the distance to the wall is not easy to measure.

\subsubsection{Swimming away from the surface}

Finally, if the swimmer starts with orientations even larger than those of the previous regime,  it will swim away from the surface. It must be remarked that, if $z(0)$ is large, even if the initial angle is negative, the reorientation produced by the interaction with the wall can be strong enough to take it away (not shown in the phase diagram of Fig. \ref{PD}). The final states in this regime are not unique and depend on the initial condition.

\subsection{Sensitivity on the gap distance}
In Section \ref{sec.regularization} it was argued the need of a small gap $\epsilon$, that would take into account the roughness at the nanometer scale and its value was fixed to $\epsilon=10\,$nm. The motion in the bulk and the swimming mode parallel to the surface are insensitive to this parameter choice not like the circular motion in contact with the surface. It is found that when changing the gap in the range $4\,$nm$<\epsilon<20\,$nm, the radius of curvature and the circular velocity vary in the range $8\,\mu$m$<R_C<50\,\mu$m and $8.6\,\mu$m/s$<V_{\rm circle}<10.8\,\mu$m/s, respectively.
This sensitivity on the value of $\epsilon$ could be responsible of the wide dispersion of velocities and radius of curvature observed in experiments~\cite{tobepublished}, due to the diversity in the nanometric roughness of the bacteria.

\section{Conclusions}
\label{sec.conclusion}
A simple model for a swimmer at zero Reynolds number has been built, consisting of two spheres of different radii joined by a dragless rod. The model aims to reproduce the swimming of a pusher bacteria like {\em E.~coli} near solid surfaces. The hydrodynamic forces and torques on each sphere are computed taking into account the full interaction with the surface using the complete resistance matrix. The hydrodynamic interaction between the spheres is treated approximately by considering the force and torque over a sphere under the flow produced by the other. The rod imposes opposite torques on the spheres making them rotate in opposite directions. Finally, the propulsion is obtained by adding an extra hydrodynamic force on one sphere (the tail) as to mimic the effect of the rotary flagella of {\em E.~coli}. The dimensions of the spheres and their separation are fixed according to experimental results of the size of  {\em E.~coli}, its velocity in the bulk, and its effective force dipole. 

When pure hydrodynamic interactions of perfect spheres are used, if the swimmer is set to approach the surface, it reaches the surface in finite time with both spheres in contact with it. At this point the swimmer stops moving due to the divergence of the hydrodynamic resistances. Appealing to the existence of nanometric roughness of the body cell or electrochemical forces with the surfaces, the resistances are regularized at the nanometric scale obtaining finite velocities when it swims close to the surface. 

The swimmer shows three different behaviors depending on the initial condition. If the swimmer is close to the surface pointing toward it, it approaches the surface and orients in such a way as to have both spheres touching the surface. Once reached this state, the swimmer performs a circular motion at constant speed with a radius of curvature in the range $8-50\, \mu$m, depending on the value used to regularize the resistances, which is in the range of the experimental observations. In the present model, the swimmer remains in this state forever. It is expected that surface roughness, thermal noise, velocity agitations or the tumbling mechanism can allow the swimmer to exit the wall region.
A second regime is obtained when the swimmer starts at higher altitudes. In this case, it approaches the surface without touching it and swimming in circles parallel to the surface with a gap of approximately a half of micron. The trajectory is also circular with a similar radius of curvature but larger velocity. To our knowledge this regime has not been identified experimentally. Finally, if the swimmer is initially far enough, it will reorient as to move away from the surface.

\section{Acknowlegment}
This research is supported by Fondecyt 1100100, Anillo ACT 127 and ECOS C07E07 grants. E.C. thanks the SESAME-Ile de France program for support.

\onecolumngrid

\appendix
\section{Fits for the resistance tensors} \label{appendix.fitsR}

Using the notation shown in Fig.~\ref{def}, the normalized distance to the wall is defined as $x=h/a$, the gap distance is $\delta=(h-a)$, and the minimum distance that the swimmer can approach to the wall is  $\epsilon$. Global fits are built such that the appropriate series when $\delta\ll a$ and when $x\gg1$ reproduce the known asymptotic expansions  (see discussion in Sec. \ref{sec.hydro}). There is not global bounding error of the fits, but we expect that the interpolating procedure produce uniformly convergent expressions if more known terms were added in both limits.
The resulting resistances for a sphere moving in a quiescent fluid and for a sphere immobilized in an ambient fluid are the following.

\subsection{Global fits for a sphere translating and rotating in a quiescent ambient fluid}

\beqa
\frac{\mathcal{R}_{\parallel}^{F U}}{6 \pi \eta a}&=&\left(2.5295 -1.9963 \, x \right)  \log \left( \frac{x-1}{x} \right)-2.9963+0.9689 \frac{1}{x}+0.5993 \frac{1}{x^2} +0.4691 \frac{1}{x^3}\\
\frac{\mathcal{R}_{\perp}^{F U}}{6 \pi \eta a}&=&-\left[2+9\left(\frac{\delta-\epsilon}{a} \right)+6\left(\frac{\delta-\epsilon}{a} \right)^2\right]\Big/\left[2 \left(\frac{\delta-\epsilon}{a} \right)+6\left(\frac{\delta-\epsilon}{a} \right)^2\right]\\
\frac{\mathcal{R}_{xy}^{F \Omega}}{6 \pi \eta a^2}&=&\left(0.3657 -0.4991 \, x \right)  \log \left( \frac{x-1}{x} \right)-0.4991+0.1162 \frac{1}{x}+0.0165 \frac{1}{x^2} -0.0028 \frac{1}{x^3}+ 0.1166 \frac{1}{x^4}\\
\frac{\mathcal{R}_{\parallel}^{\tau \Omega}}{8 \pi \eta a^3}&=&\left(-0.3898 +0.7898 \, x \right)  \log \left( \frac{x-1}{x} \right)-0.2101+0.0050 \frac{1}{x}+0.0683 \frac{1}{x^2} -0.2449 \frac{1}{x^3}\\
\frac{\mathcal{R}_{\perp}^{\tau \Omega}}{8 \pi \eta a^3}&=&-\left(0.414 (x-1) +0.318 \, \frac{(x-1)^2}{x}\right)  \log \left( \frac{x-1}{x} \right)-1.732 + 0.684 \frac{1}{x}-0.037 \frac{1}{x^2} -0.117 \frac{1}{x^3}
\eeqa

\subsection{Global fits for a sphere immobilized in an ambient fluid}
\beqa
\frac{\mathcal{R}_{\parallel}^{F U^\infty}}{6 \pi \eta a}&=&1+\frac{9}{16}\left(\frac{a}{h} \right)+0.1375\left(\frac{a}{h} \right)^2\\
\frac{\mathcal{R}_{\perp}^{F U^\infty}}{6 \pi \eta a}&=&1+\frac{9}{8}\left(\frac{a}{h} \right)\\
\frac{\mathcal{R}_{\parallel}^{\tau \Omega^\infty}}{8 \pi \eta a^3}&=&1-\frac{3}{16}\left(\frac{a}{h} \right)^3+0.1315\left(\frac{a}{h} \right)^4\\
\frac{\mathcal{R}_{\perp}^{\tau \Omega^\infty}}{8 \pi \eta a^3}&=&1+\frac{1}{8}\left(\frac{a}{h} \right)^3\\
\mathcal{R}^{F \Omega^\infty}&=&\mathcal{R}^{\tau U^\infty}=0
\eeqa

\twocolumngrid

\end{document}